\documentclass{aastex631}

\usepackage{amsmath}
\usepackage{CJK}

\submitjournal{ApJ}

\shorttitle{Probing time variation of the fine structure constant}
\shortauthors{Zhi-E Liu et al.}

\graphicspath{{./}{figures/}}

\begin{document}
\begin{CJK*}{GB}{gbsn}

\title{Probing the time variation of fine structure constant using galaxy clusters and quintessence model}


\correspondingauthor{Zhi-E Liu(ÁõÖ¾¶ð), Tong-Jie Zhang(ÕÅͬ½Ü)}
\email{zhieliu@163.com, tjzhang@bnu.edu.cn}

\author[0000-0003-3153-1296]{Zhi-E Liu(ÁõÖ¾¶ð)}
\affiliation{College of Physics and Electronic Engineering, Qilu Normal University \\
 Jinan 250200, China}

\author{Wen-Fei Liu(ÁõÎÄì³)}
\affiliation{College of Physics and Electronic Engineering, Qilu Normal University \\
 Jinan 250200, China}

\author[0000-0002-3363-9965]{Tong-Jie Zhang(ÕÅͬ½Ü)}
\affiliation{College of Physics and Electronic Engineering, Qilu Normal University \\
 Jinan 250200, China}
\affiliation{Department of Astronomy, Beijing Normal University \\
Beijing 100875, China}

\author{Zhong-Xu Zhai(µÔÖÒÐñ)}
\affiliation{IPAC, California Institute of Technology \\
Mail Code 314-6, 1200 E. California Blvd., Pasadena, CA 91125}


\author[0000-0002-6372-9363]{Kamal Bora}
\affiliation{Department of Physics, Indian Institute of Technology \\
Hyderabad, Kandi, Telangana-502284, India}

\begin{abstract}
We explore a possible time variation of the fine structure constant ($\alpha \equiv e^2/\hbar c$) using the Sunyaev-Zel'dovich effect measurements of galaxy clusters along with their X-ray observations. Specifically, the ratio of the integrated Compto-ionization parameter $Y_{SZ}D_A^2$ and its X-ray counterpart $Y_X$ is used as an observable to constrain the bounds on the variation of $\alpha$. Considering the violation of cosmic distance duality relation, this ratio depends on the fine structure constant as $\sim \alpha^3$. We use the quintessence model to provide the origin of $\alpha$ time variation. In order to give a robust test on $\alpha$ variation, two galaxy cluster samples, the 61 clusters provided by the Planck collaboration and the 58 clusters detected by the South Pole Telescope, are collected for analysis. Their X-ray observations are given by the XMM-Newton survey. Our results give $\zeta=-0.203^{+0.101}_{-0.099}$ for the Planck sample and $\zeta=-0.043^{+0.165}_{-0.148}$ for the SPT sample, indicating that $\alpha$ is constant with redshift within $3\sigma$ and $1\sigma$ for the two samples, respectively.

\end{abstract}


\section{Introduction} \label{sec:intro}
The fine structure constant $\alpha \equiv e^2/\hbar c$ is at the central position in the system of fundamental physical constants. It measures the strength of the electromagnetic interaction between charged elementary particles in the low-energy limit. Recently, the fine structure constant was determined at an unprecedented precise, that is, $\alpha^{-1}=137.035999206$ with a relative accuracy of $81$ parts per trillion \citep{Morel2020}. However, in 1937, Dirac argued that the
fundamental constants of Nature may not be pure constants but vary slowly with the epoch, and he proposed a gravitational `constant' decreasing proportionally to $t^{-1}$\citep{Dirac1937}. Since then, some theoretical and experimental investigations allowing space-time variation of fundamental constants have been put into effect \citep{Martins2017,Uzan2011,Uzan2003,Wang2020}. In order to remedy the dire consequence \citep{Teller1948} induced by the varying gravitational constant $G \sim t^{-1}$, Gamow(in 1967) suggested that $e^2$ increases in direct proportion to the age of the universe \citep{Gamov1967}. Phenomenological models that usually assumed an $\alpha$
varying as some power-law or logarithm of time, like those introduced by Gamow, represent the early work of studying varying $\alpha$ \citep{Barrow1986}.

Over the last few decades, the focus has shifted to various extensions of standard models, in which one or more fundamental constants of Nature become dynamical quantities. The first self-consistent theory of varying $\alpha$ is the framework constructed by \citet{Bekenstein1982} via modifying Maxwell electrodynamics, which was then extended to a cosmological setting by \citet{Sandvik2002,Barrow2002a,Barrow2003,Barrow2002b}, namely BSBM theory. In the BSBM model, variations in $\alpha$ occur due to a coupling between the electromagnetic field and a massless scalar field in the action. The original BSBM model was then extended by adding a non-trivial potential \citep{Barrow2008,Farajollahi2012}, or by allowing the coupling between scalar field $\phi$ and photons as a function of $\phi$ \citep{Barrow2012}, or by allowing for both an arbitrary coupling and potential function \citep{Barrow2013}. Another class of extensions of standard models are based on the belief that space has more than three dimensions, and the extra space dimensions can cause varying fundamental constants \citep{Chodos1980,Marciano1984,Kolb1986,Barrow1987}. For instance, in Kaluza-Klein theories, the time evolution of the mean KK radius $R_{\textrm{KK}}$ of extra spatial dimensions can give rise to time variations of fundamental constants \citep{Marciano1984}.

At present, observational data of $\alpha$ variation are becoming increasingly abundant. The cosmic microwave background (CMB) data \citep{Bryan2015,Planck2015,Hart2018} and the Big Bang nucleosynthesis (BBN) \citep{Iocco2009,Mosquera2013,Clara2020} can constrain $\alpha$ variations in the early universe. The Planck Collaboration obtained $\Delta \alpha/\alpha\approx 10^{-3}$ at redshift $z\approx 10^3$ by analyzing the CMB spectra \citep{Planck2015}, and the constraints given by the abundance of the light elements emerged during BBN are not very tight (approximately $\Delta \alpha / \alpha < 10^{-2}-10^{-3}$ at $z\approx 10^9-10^{10}$), too \citep{Mosquera2013}. The 1.8 billion-year-old natural nuclear reactor at the Oklo Uranium Mine in Gabon can give much tighter bounds on the time variations of $\alpha$ ($\Delta \alpha / \alpha \approx 10^{-7}-10^{-8}$) \citep{Damour1996,Lamoreaux2004}. The most sensitive constraints on $\Delta \alpha / \alpha$ were achieved at $z \approx 1-6$ from the spectral lines significantly affected by relativistic effects in absorbing clouds around distant quasars. The quasar absorption lines observed by Keck/HIRES and VLT/UVES telescopes have produced a large data sample that describe the dependence of $\Delta \alpha / \alpha$ on redshift $z$, in which the variations of $\alpha$ are normally constrained by $\Delta \alpha / \alpha \approx 10^{-5}-10^{-6}$ \citep{Murphy2001a,Murphy2001b,Murphy2001c,Murphy2003,Murphy2004,Murphy2007,Murphy2008,Chand2004,Srianand2004,Webb1999,Webb2001}. In addition, the spatial variations of $\alpha$ observed by comparing the results from Keck and VLT have also got interest of study. Specifically, at $z>1.8$ the Northern sky observations of the Keck telescope suggested a smaller value of the fine structure constant ($\Delta \alpha / \alpha=(-0.74\pm 0.17)\times 10^{-5}$), but the Southern sky observations of the VLT telescope showed an increasing fine structure constant ($\Delta \alpha / \alpha=(0.61\pm 0.20)\times 10^{-5}$) \citep{Webb2011,King2012}. This could be due to an undetected systematic effects, but may also hint to a new physics. More recently, a constraint on the relative variation of $\alpha$ below $10^{-5}$ is obtained by comparing the absorption lines of late-type evolved giant stars from the S star cluster orbiting the supermassive black hole in our Galactic Center with the absorption lines in the lab \citep{Hees2020}. The absorption lines of white dwarf stars can also be used to constrain $\Delta \alpha / \alpha$ \citep{Berengut2013}.

Some work considers a class of dilaton runaway models, where $\Delta\alpha / \alpha = -\gamma \ln (1+z)$, and indirectly constrains $\Delta\alpha / \alpha$ by constraining $\gamma$. Most of this kind of work employed the observations of the Sunyaev-Zeldovich effect combined with observations of the X-ray surface brightness of galaxy clusters, for which the former can be characterized by the integrated Compto-ionization parameter $Y_{SZ}D^2_A$ and the latter by the $Y_X$
parameter \citep{Holanda2017,Bora2021a,Colaco2019,Holanda2016a,Holanda2016b,Bora2021b}. Alternatively, Cola\c{c}o et al. used combined measurements of Strong Gravitational Lensing systems and Type Ia Supernovae to constrain $\gamma$ \citep{Colaco2020a,Colaco2020b}. Instead of based on runaway models, Galli assumed that $\alpha$ can linearly vary with redshift, i.e., $\alpha / \alpha_0 = A_{lin}(1+z)$, and then studied whether $\alpha$ is time-dependent by constraining $A_{lin}$ using the Sunyaev-Zeldovich effect and its X-ray counterpart of galaxy clusters \citep{Galli2013}.

In this paper, we will consider a specific model for the variation of the fine structure constant $\alpha$ driven by a typical quintessence scenario, i.e. a linear coupling with the electromagnetic term in the Lagrangian.

\section{COUPLING OF QUINTESSENCE TO THE ELECTROMAGNETIC FIELD}
 Quintessence is one type of dynamical scalar field models 
 \citep{Padmanabhan2003}. It has been used to model the dark energy component in the universe. We expect the quintessence field to couple with the electromagnetic sector of the matter-radiation Lagrangian and induce a time variation of the fine structure constant $\alpha$. The general form of the action involving a quintessence scalar field and its coupling with the electromagnetic term can be written as \citep{Marra2005}
 \begin{equation}
 \begin{split}
 S=&\frac{1}{16\pi G}\int \textrm{d}^4x\sqrt{-g}R + \int \textrm{d}^4x\sqrt{-g}\left[\frac{1}{2}\partial^\mu\phi\partial_\mu\phi-V(\phi)\right] \\
 - &\frac{1}{4}\int \textrm{d}^4x\sqrt{-g}B_F(\phi)F_{\mu\nu}F^{\mu\nu} + \int \textrm{d}^4x\sqrt{-g} \mathcal{L}
 \end{split}
 \label{eq1}
  \end{equation}
where $G$ is the Newton's Gravitational constant and $\mathcal{L}$ represents the Lagrangian density including the Standard Model fields and a hypothetical dark matter sector. $B_F(\phi)$ describes the coupling between quintessence and the electromagnetic field and allows for the evolution in $\phi$.  The effective fine structure constant depends on the value of $\phi$ as \citep{Copeland2004}
\begin{equation}
 \alpha(\phi) = \frac{\alpha_0}{B_F(\phi)}.
 \label{eq2}
\end{equation}
where $\alpha_0$ is the fine structure constant measured today. Therefore, we get the relative variation of $\alpha$:
\begin{equation}
 \frac{\Delta\alpha}{\alpha} \equiv \frac{\alpha(t)-\alpha_0}{\alpha_0} = \frac{1-B_F(\phi)}{B_F(\phi)}.
 \label{eq3}
\end{equation}

We consider a homogeneous and isotropic FRW cosmology described by the line element
\begin{equation}
 \textrm{d}s^2 = -\textrm{d}t^2 + a^2(t)\left[\frac{\textrm{d}r^2}{1-Kr^2}+r^2(\textrm{d}\theta^2+\sin^2\theta \textrm{d}\varphi^2)\right]
 \label{eq4}
\end{equation}
where $a(t)$ and $K$ are the scale factor and spatial curvature respectively. Combining Eq.(\ref{eq1}) and Eq.(\ref{eq4}) can yield equations describing the evolution of the quintessence scalar field in the FRW universe, that is, the Friedmann equation \citep{Park2019}
\begin{equation}
 \left(\frac{H}{H_0}\right)^2 = \frac{1}{1-\frac{1}{6}(\phi')^2}\left[\Omega_m a^{-3}+\Omega_r a^{-4}+ \Omega_k a^{-2} + \frac{V(\phi)}{3H^2_0} \right]
 \label{eq5}
\end{equation}
and the Klein-Gordon equation \citep{Park2019}
\begin{equation}
 \phi'' + \left(3+\frac{\dot{H}}{H^2}\right)\phi' + \frac{dV(\phi)}{d\phi}\frac{1}{H^2} = 0,
 \label{eq6}
\end{equation}
where $\phi' \equiv d\phi/d\ln a$, and an overdot denotes the time derivative $d/dt$. We have set the present scale factor $a_0=1$ and chosen the units such that the Newtonian gravitational constant $G \equiv 1/8\pi$. $H=\dot{a}/a$ is the Hubble parameter and $H_0$ is the Hubble constant. The present value of the non-relativistic matter density parameter $\Omega_m$ is the sum of present baryonic matter and CDM density parameters, $\Omega_m=\Omega_b+\Omega_c$, $\Omega_r$ is the present value of the radiation density parameter, and $\Omega_k$ is the present value of the spatial curvature density parameter. We assume a spatially flat universe, $\Omega_k=0$, and set $\Omega_m=0.315\pm 0.007$, $H_0=67.4 \pm 0.5 \textrm{km s}^{-1} \textrm{Mpc}^{-1}$ according to the Planck 2018 results \citep{Planck2020,Chen2019}. $\Omega_r$ is not a free parameter but determined by $\Omega_r=\Omega_m a_{eq}$, where $a_{eq}$ is the scale factor at the epoch of matter-radiation equality given by $a_{eq} = \frac{4.15 \times 10^{-5}}{\Omega_m h^2}$ \citep{Dodelson2020}.
Here $h=H_0/(100 \textrm{km s}^{-1} \textrm{Mpc}^{-1})$.

A general functional form of the quintessence potential $V(\phi)$ can be written as a combination of power-law and exponential functions. \citet{Copeland2004} discussed in detail the effect induced by different quintessence models on the cosmological $\Delta \alpha$. In our work, we only consider an inverse power-law potential \citep{Peebles1988,Samushia2006,Samushia2009,Chen2011,Park2019}
\begin{equation}
 V(\phi) = V_1\phi^{-n}
 \label{eq8}
\end{equation}
where $V_1$ and $n$ are non-negative constant parameters. In the limit $n=0$ the quintessence dark energy becomes the cosmological constant
$\Lambda$. The exponent parameter $n$ has been constrained to $[0,6]$ by various observational data \citep{Park2019}. \citet{Marra2005} showed that choosing different quintessence potentials gives a subdominant effect on the cosmological variation of $\alpha$ with respect to changing the coupling function $B_F(\phi)$. Our tests also show that different values of $n$ only have a mild impact on the result, and fixing this parameter can improve the efficiency of the computation significantly. Thus, following \citet{Marra2005} we set $n=1.0$ in order to have the correct attractor equation of state \citep{Marra2005}. By definition, the pressure $p_{\phi}$ and energy density $\rho_{\phi}$ of the scalar field are given by
\begin{equation}
 \begin{split}
 p_{\phi} = \frac{\dot{\phi}^2}{2} - V(\phi), \\
 \rho_{\phi} = \frac{\dot{\phi}^2}{2} + V(\phi),
 \end{split}
 \label{eq9}
\end{equation}
and the equation of state for scalar field is $\omega_{\phi}=p_{\phi}/\rho_{\phi}$.
The coefficient parameter $V_1$ of the potential (\ref{eq8}) can be numerically calculated by making the energy density today $\rho^{(0)}_{\phi}$, which is evolved from equation (\ref{eq6}), equal to
\begin{equation}
\rho_{cr}\Omega_{\phi} = \rho_{cr}(1-\Omega_m-\Omega_r-\Omega_k),
 \label{eq10}
\end{equation}
where $\rho_{cr}=3H_0^2$ is the critical density. Thus, $V_1$ is in fact determined by $n$ implicitly. Accordingly, the dark energy density parameter today is $\Omega_{\phi} = \rho_{\phi}^{(0)}/\rho_{cr}=(\phi'_0)^2/6 + V(\phi_0)/(3H_0^2)$, where $\phi_0$ and $\phi'_0$ are the current values of $\phi$ and $\phi'$.

We numerically solve the cosmological equations (\ref{eq5})-(\ref{eq6}) and then produce the resulting $\Delta \alpha$ at a series of redshifts. We use the initial conditions \citep{Peebles1988}
\begin{equation}
\phi = \left[\frac{2}{3}n(n+2)\right]^{1/2}\left(\frac{a}{a_1}\right)^{3/(n+2)}
 \label{eq11}
\end{equation}
at scale factor $a_i=10^{-10}$. Here $a_1$ denotes the characteristic epoch at which the dominant mass density switches from ordinary matter to the $\phi$ field.

Equation (\ref{eq3}) shows the evolution of $\alpha$ depends on the evolution of $\phi$ via a concrete functional form of $B_F(\phi)$, so what follows is the choice of $B_F(\phi)$. In principle, there are no constraints on the form of $B_F(\phi)$. \citet{Marra2005} proposed a general form of the function $B_F(\phi)$ which is characterized by a set of four parameters, and comprehensively discussed various $B_F(\phi)$ cases that gave different $\alpha$ histories. In this work we adopt the simplest case that is originally proposed by \citet{Bekenstein1982}, i.e. a linear dependence on $\phi$ such that
\begin{equation}
 B_F(\phi) = 1-\zeta(\phi-\phi_0)
 \label{eq12}
\end{equation}
where $\phi_0$ denotes the present quintessence, and $\zeta$ is a parameter describing the strength of the coupling between quintessence and the electromagnetic field and to be determined by the observational data. The case $\zeta=0$, or $B_F(\phi)=1$, means there is no coupling to the electromagnetic field, and thus the fine structure constant $\alpha$ really stays a constant during the lifetime of the universe. Substituting Eq.(\ref{eq12}) into Eq.(\ref{eq3}) results in
\begin{equation}
 \frac{\Delta\alpha}{\alpha} = \zeta(\phi-\phi_0).
 \label{eq13}
\end{equation}

\section{$Y_{\textrm{SZ}}$ - $Y_X$ relation and $\alpha$}
The Sunyaev-Zel¡¯dovich Effect ($Y_{SZ}$) and X-ray surface brightness ($Y_X$) are two observable quantities of galaxy clusters. Since $Y_{SZ}$ and $Y_X$ both reflects the
thermal energy of the cluster and are proportional to the total cluster mass, their ratio $Y_{SZ}D_A^2/Y_X$ should be a constant independent of redshift. This ratio has been used to explore the variation of $\alpha$ \citep{Galli2013,Colaco2019,Holanda2016b,Bora2021a,Bora2021b}, which is also our main focus in this paper.

The high energy electrons in the ionized intergalactic gas of galaxy clusters can scatter the CMB photons, via Inverse Compton Effect, resulting in the distortion of the CMB spectrum. This is the known Sunyaev-Zel'dovich effect (SZ, hereafter) \citep{Sunyaev1972}. The CMB spectral distortion caused by SZ effect is proportional to the Compton parameter $y$, which is expressed as
 \begin{equation}
 y=\frac{\sigma_T k_B}{m_e c^2}\int n_e Tdl = \frac{\sigma_T}{m_e c^2}\int Pdl
 \label{eq14}
\end{equation}
where $k_B$ is the Boltzmann constant, $c$ is the speed of light, $m_e$ is the electron mass, $n_e$ is the number density of electrons, $T$ is the electron temperature, and $P=n_e k_B T$ is the pressure of the intracluster medium under the assumption of ideal gas equation of state. Therefore, the Compton parameter quantifies the gas pressure of the intracluster medium integrated along the line of sight. The Thompson cross-section $\sigma_T$ can be computed using Feynman diagrams and linked to the fine structure constant $\alpha$ by
 \begin{equation}
 \sigma_T = \frac{8\pi}{3}\left(\frac{e^2}{m_ec^2}\right)^2 = \frac{8\pi}{3}\left(\frac{\hbar^2 \alpha^2}{m_e^2 c^2}\right)
 \label{eq15}
\end{equation}
Integrating the Compton parameter $y$ over the plane perpendicular to the line of sight can provide the spherical integrated Compton parameter, defined as
 \begin{equation}
 Y_{SZ}D_A^2 = \int ydA = \frac{\sigma_T k_B}{m_e c^2}\int n_e T dl dA = \frac{\sigma_T}{m_e c^2}\int PdV.
 \label{eq16}
\end{equation}
Thus, we can see that $Y_{SZ}D_A^2$ depends on the fine structure constant through the Thompson cross-section (see Eq.(\ref{eq15})) as \citep{Galli2013}
\begin{equation}
 Y_{SZ}D_A^2 \propto \alpha^2.
 \label{eq17}
\end{equation}
According to Eq.(\ref{eq2}), we have
\begin{equation}
 Y_{SZ}D_A^2 \propto B_F^{-2}(\phi).
 \label{eq18}
\end{equation}

The intergalactic hot gas emits mainly through thermal bremsstrahlung. The magnitude of the X-ray emission is quantified by the $Y_X$ parameter which can be acquired by X-ray surface brightness observations and is defined as \citep{Kravtsov2006}
\begin{equation}
 Y_X = M_g(R)T_X,
 \label{eq19}
\end{equation}
where $M_g(R)$ is the X-ray determined gas mass within the radius $R$ and $T_X$ is the X-ray temperature of the cluster. It has been shown that $M_g(R)$ can be
written as \citep{Goncalves2012,Colaco2019}
\begin{equation}
 M_g(<\theta)\propto \alpha^{-3/2} D_L D_A^{3/2}
 \label{eq20}
\end{equation}
where $D_L$ is the luminosity distance. This equation shows $M_g(R)$ depends not only on the fine structure constant but also on the validity of the cosmic distance duality relation (CDDR), $D_L=(1+z)^2D_A$. A variation of $\alpha$ will lead to a violation of the CDDR \citep{Hees2014}, which is normally described as $D_L = \eta(z)(1 + z)^2D_A$. Consequently, $Y_X$ will scale to $\alpha(z)$ and $\eta(z)$ by
 \begin{equation}
 Y_X \propto M_g(<\theta) \propto \alpha(z)^{-3/2} \eta(z)
 \label{eq21}
\end{equation}
According to the scalar field theory, $\alpha(z)$ and $\eta(z)$ are linked by \citep{Hees2014}
 \begin{equation}
 \frac{\Delta\alpha}{\alpha} \equiv \frac{\alpha(z)-\alpha_0}{\alpha_0} = \eta(z)^2-1.
 \label{eq22}
\end{equation}
Together with Eq.(\ref{eq2}) we have
\begin{equation}
Y_X \propto \alpha^{-1} \propto B_F(\phi).
\label{eq23}
\end{equation}
 Then, based on Eqs.(\ref{eq18}) and (\ref{eq23}), the dimensionless ratio of $Y_{SZ}D_A^2$ to $Y_X$ can be related to the coupling strength $B_F(\phi)$ by
\begin{equation}
\frac{Y_{SZ}D_A^2}{Y_X} \propto B_F(\phi)^{-3}.
\label{eq24}
\end{equation}
This ratio can also been expressed in the following form \citep{Galli2013}:
\begin{equation}
\frac{Y_{SZ}D_A^2}{Y_X} \propto C_{XSZ}\frac{\int n_e(r)T(r)dV}{T(R)\int n_e(r) dV},
\label{eq25}
\end{equation}
where
\begin{equation}
C_{XSZ} = \frac{\sigma_T}{m_ec^2}\frac{1}{\mu_em_p} \approx 1.416 \times 10^{-19} \left(\frac{Mpc^2}{M_{\odot}keV}\right).
\label{eq26}
\end{equation}
The numerator and denominator in Eq.(\ref{eq25}) are both approximations of the thermal energy of the cluster. As discussed in \citet{Colaco2019}, this ratio would be exactly constant with redshift if no new physics is assumed. As done by \citet{Galli2013,Colaco2019,Bora2021a}, and considering Eq.(\ref{eq12}), we can rewrite the ratio in Eq.(\ref{eq24}) into
\begin{equation}
\frac{Y_{SZ}D_A^2}{Y_XC_{XSZ}} = CB_F(\phi)^{-3} = C(1-\zeta(\phi-\phi_0))^{-3},
\label{eq27}
\end{equation}
where $C$ is a constant to be determined. The case of $C \simeq 1$ indicates that the galaxy clusters used in this analysis are isothermal.

We are interested in a possible $\alpha$ variation as predicted by the quintessence model. The quintessence scalar field $\phi$ evolves via Eqs.(\ref{eq5}) and (\ref{eq6}), then the the constants $C$ and $\zeta$ can be determined by observational measurements. As shown in Eq.(\ref{eq13}), if the resulting $\zeta$ departs from zero significantly, a time variation for $\alpha$ should be established.

\section{Galaxy cluster samples}
Large solid angle surveys employing the SZ effect have been carried out with the South Pole Telescope (SPT, \citet{Carlstrom2011}), Planck \citep{Planck2011b}, and the Atacama Cosmology Telescope (ACT, \citet{Fowler2007}). One data set used here is the SZ and X-ray data extracted from Table 1 of \citet{Planck2011b}. The SZ effect measurements in the direction of 62 nearby galaxy clusters ($z < 0.5$), which is a sub-sample of the Planck Early Sunyaev-Zel'dovich (ESZ) cluster sample \citep{Planck2011a}, were detected at high signal-to-noise ($S/N \geq 6$) in the first Planck all-sky data set. These galaxy clusters are not contaminated by flares and their morphologies are regular enough that the spherical symmetry can be assumed. They had also been observed by the XMM-Newton telescope, and their $Y_X$ measurements were obtained with the deep XMM-Newton X-ray data \citep{Piffaretti2011}. As done by \citet{Galli2013}, cluster A2034 is excluded from the analysis since its redshift estimate is discordant in \citet{Planck2011b,Mantz2010}. So we use 61 Planck-ESZ clusters in this analysis covering the redshift range $0.044<z<0.44$. The $Y_{SZ}$ and $Y_X$ parameters are labeled as $D_A^2Y_{500}$ and $Y_{X,500}$ respectively. Here 500 indicates a radius $R_{500}$ at which the mean matter density of the cluster is 500 times than the critical density of the Universe at the redshift of the cluster. The computation of the $Y_{SZ}$ parameter requires shaping the thermal pressure ($P$) of the intracluster medium for each galaxy cluster by using the Universal Pressure Profile \citep{Arnaud2010}. Furthermore, the X-ray temperatures are defined within a cylindrical annulus of radius $[0.15-0.75]R_{500}$. See \citet{Galli2013,Colaco2019} for more comments on the data set.

Another data set used in our analysis is the SPT galaxy cluster sample. SPT is a 10m telescope at the South Pole that images the sky at three different frequencies, i.e. 95 GHz, 150 GHz, and 220 GHz \citep{Carlstrom2011}. SPT has detected 516 galaxy clusters via the SZ effect in the 2500 square degree SPT-SZ Survey at $0<z<1.8$ with masses $M_{500}\geq3\times 10^{14}M_{\bigodot}$, and approximately 20 percent of the sample lies at $z > 0.8$ \citep{Bleem2015}. The redshifts of the SPT cluster candidates have been obtained through dedicated optical surveys and follow-up programs \citep{Desai2012,Song2012,Saro2015}. Their properties were described in \citet{Bleem2015}, among which the redshift values were updated in \citet{Bocquet2019}. The $Y_{SZ}$ value of each SPT cluster is measured through integrating the thermal SZ signal in a cylindrical volume within a $0.75'$-radius aperture. Note their X-ray counterpart, $Y_X$, is measured within a 3D sphere of radius $R_{500}$. In order to compare $Y_{SZ}$ with $Y_{X}$, one needs to convert the SPT $Y_{SZ}$ values to spherically integrated $Y_{SZ}$ within the same radius at which $Y_X$ was measured. This conversion is performed based on the following expressions \citep{Arnaud2010}:
\begin{equation}
Y_{\textrm{cyl}}(R_1) = Y_{\textrm{sph}}(R_b) - \frac{\sigma_T}{m_ec^2} \int_{R_1}^{R_b} 4\pi P(r)\sqrt{r^2-R_1^2}r\textrm{d}r
\label{eq28}
\end{equation}
\begin{equation}
Y_{\textrm{sph}}(R_2) =  \frac{\sigma_T}{m_ec^2} \int_{0}^{R_2} 4\pi P(r)r^2\textrm{d}r.
\label{eq29}
\end{equation}
 Here, $Y_{\textrm{cyl}}(R_1)$ is the $Y_{SZ}$ parameter measured within a cylindrical aperture of radius $R_1$, and $Y_{\textrm{sph}}(R_2)$ is the $Y_{SZ}$ parameter integrated within a spherical volume of radius $R_2$. $R_b$ is the radial extent of the cluster, and $P(r)$ is the gas pressure in the intra-cluster medium. An analytical form of $P(r)$ is needed to calculate the integrations in Eqs.(\ref{eq28}-\ref{eq29}), for which we adopt the Universal Pressure Profile suggested by \citet{Arnaud2010}. Following \citet{Bora2021a}, we set $R_b = 10R_{500}$. Since the SPT observation gave $Y_{SZ}$ values in terms of an aperture of $0.75'$, $R_1 = 0.75'D_A$ is required in Eq.(\ref{eq28}). We put $R_2 = R_{500}$ in Eq.(\ref{eq29}), concordant with the radius at which the SPT $Y_X$ is measured. In order to be less dependent on the free parameters involved in the Universal Pressure Profile of $P(r)$ and make the computation more reliable, instead of calculating $Y_{\textrm{sph}}(R_{500})$ directly by Eq.(\ref{eq29}), we estimate $Y_{\textrm{sph}}(R_{500})$ via calculating the ratio of $Y_{\textrm{sph}}(R_{500}$) to $Y_{\textrm{cyl}}(0.75'D_A)$ from Eqs.(\ref{eq28}) and (\ref{eq29}).

 The XMM-Newton X-ray observations of 73 of these SZE-selected clusters have been performed by SPT collaboration members or various non-SPT small programs, among which 15 clusters were excluded due to their low data quality or low redshift ($z<0.2$). Thus, a sample of 58 clusters with redshift range $0.2 < z < 1.5$ is used for analysis. The sample has a median mass and redshift of $M_{500}=4.77 \times 10^{14}M_{\bigodot}$ and $z_{med} = 0.45$ respectively, and five clusters lie at $z > 1$.The X-ray observable-mass scaling relations of these clusters were studied in \citet{Bulbul2019}, and the details of their XMM-Newton observations, especially $Y_X$ values, can be found in Table 1 of \citet{Bulbul2019}. Same as the Planck counterparts, the X-ray observables of the sample were measured at $R_{500}$.
 \begin{figure}
 \fig{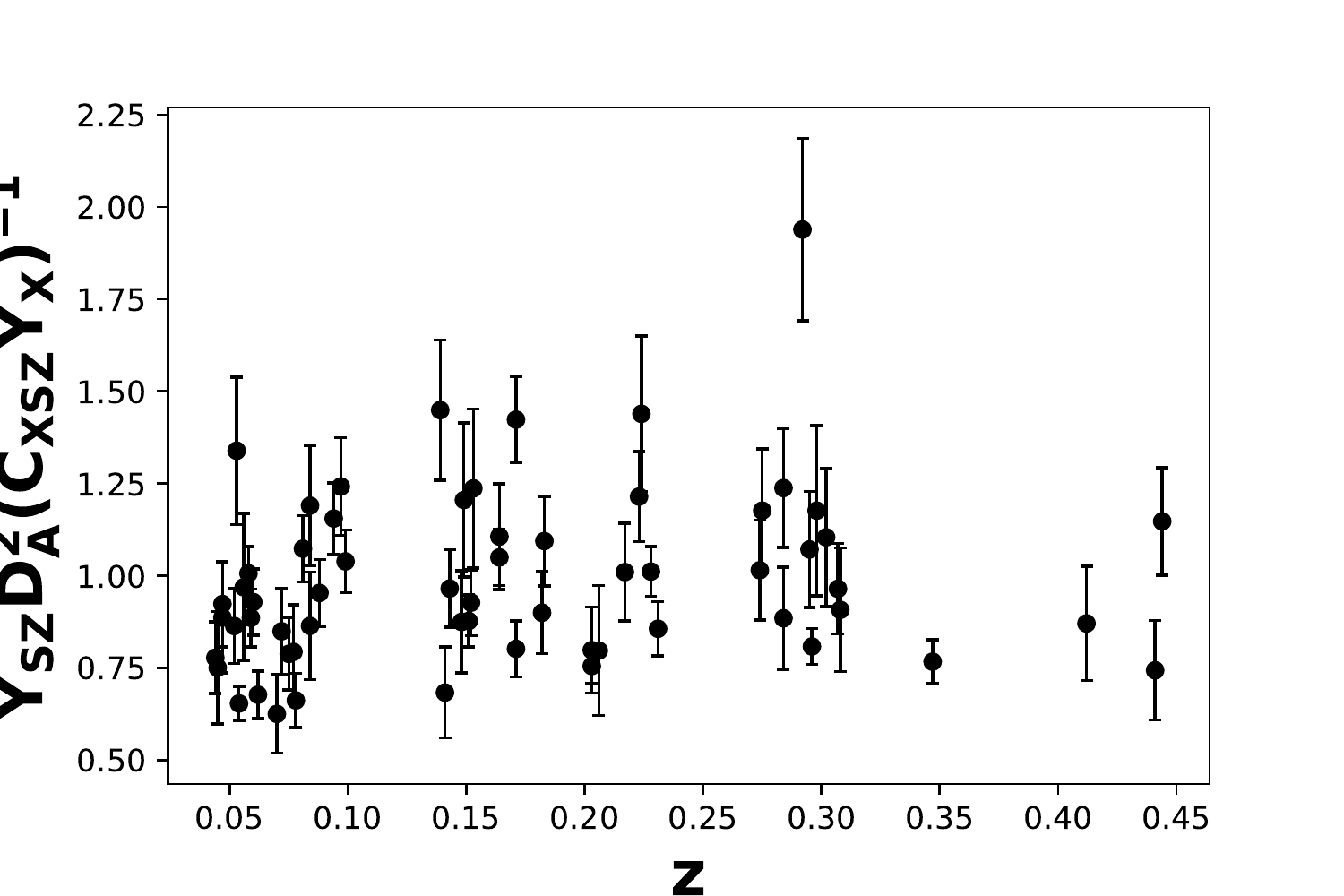}{0.48\textwidth}{(a)}
 \fig{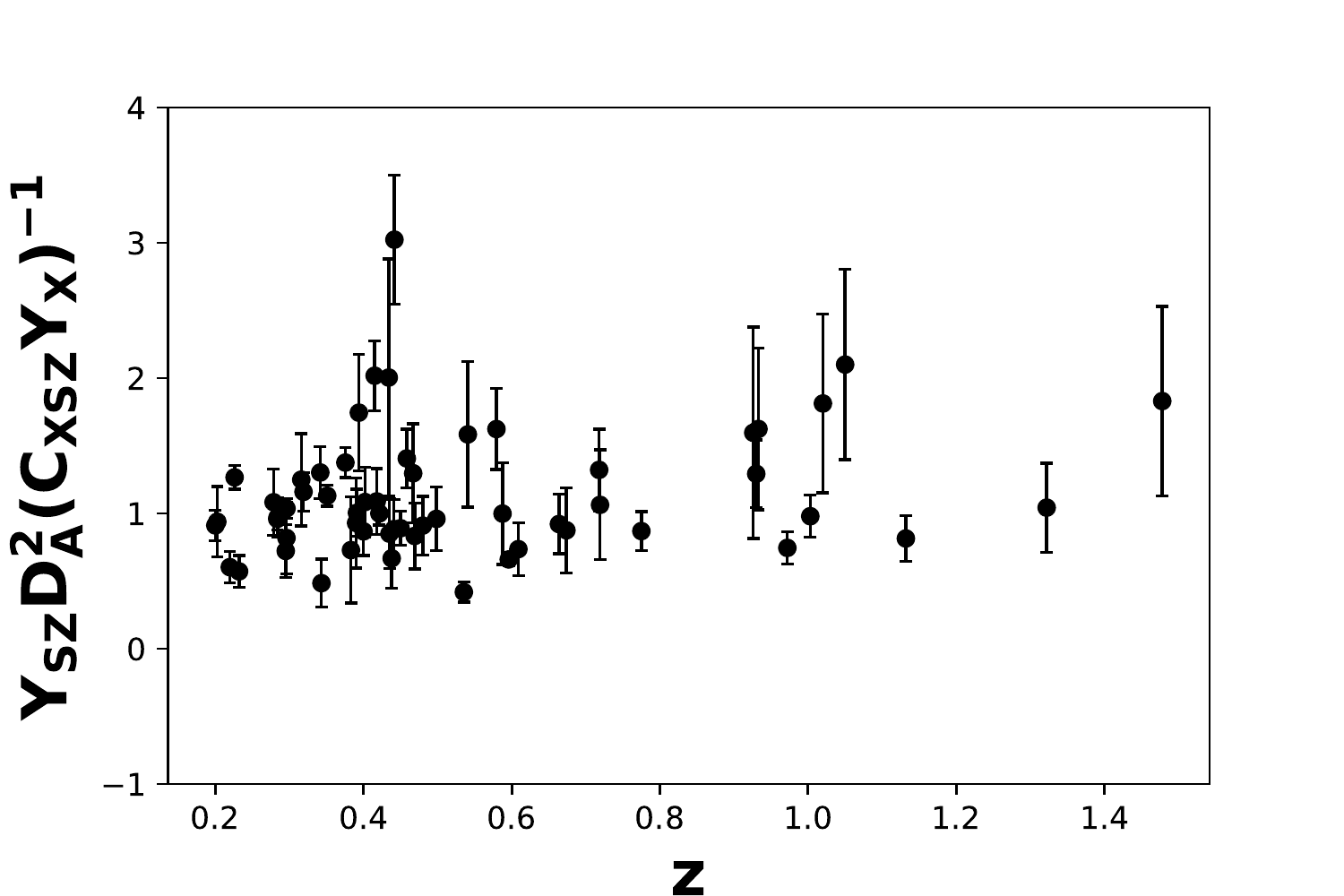}{0.48\textwidth}{(b)}
 \caption{The observed ratio $Y_{SZ}D^2_A/Y_XC_{XSZ}$. They are calculated from (a) the Planck galaxy cluster sample, (b) the SPT galaxy cluster sample.}
  \label{fig1}
\end{figure}

\section{Analysis and Results}
In order to constrain the parameters of Eq.~\ref{eq27}, we minimize the following negative log-likelihood function:
\begin{equation}
-2\ln \textsl{L} = \sum_{i=1}^{N} \ln 2\pi \sigma_i^2 + \sum_{i=1}^{N} \frac{[R_{obs,i}- C(1-\zeta(\phi-\phi_0))^{-3}]^2}{\sigma_i^2},
\label{eq30}
\end{equation}
where
\begin{equation}
R_{obs,i} = \frac{(Y_{SZ}D_A^2)_i}{Y_{X,i}C_{XSZ}}.
\label{eq31}
\end{equation}
denotes the observed values of the ratio in Eq.(\ref{eq27}) for galaxy clusters, $N$ is the total number of clusters, and $\sigma_i$ is the total uncertainty inherited from the observations $(Y_{SZ}D^2_A)_i$ and $Y_{X,i}$, plus an intrinsic scatter term $\sigma_{int}$,  that is
\begin{equation}
\sigma_i^2 = \left(\frac{\sigma_{SZ,i}}{Y_{X,i}C_{XSZ}}\right)^2 + \left(\frac{(Y_{SZ}D_A^2)_i\sigma_{X,i}}{Y_{X,i}C_{XSZ}}\right)^2 +\sigma_{int}^2.
\label{eq32}
\end{equation}
where $\sigma_{SZ,i}$ and $\sigma_{X,i}$ denote the uncertainty of $(Y_{SZ}D_A^2)_i$ and $Y_{X,i}$, respectively. The advantage of using log-likelihood against the commonly used $\chi^2$ criterion is that the log-likelihood function can accommodate the optimization of the intrinsic uncertainty. Thus, for the SPT sample the parameter space to be optimized is $\{C,\zeta,\sigma_{int}\}$. For the Planck sample, we followed \citet{Galli2013} and set $\sigma_{int} = 0.17$. So the parameter space of the Planck sample is reduced to $\{C,\zeta\}$, and the objective function (\ref{eq30}) becomes equivalent to the $\chi^2$ criterion. Our experiment shows different cosmological parameters only have marginal effect on the result, so we fix the cosmological parameters as $H_0=67.4$, $\Omega_m=0.315$ in the implementation of quintessence model.

Unlike the Planck cluster sample for which the product of $Y_{SZ}$ and $D_A^2$ i.e. $Y_{SZ}D_A^2$, is already given as an  observational quantity, for SPT data we need to estimate the angular diameter distance to each galaxy cluster via,
\begin{equation}
D_A = \frac{1}{1+z}\int _0^z \frac{c dz'}{H(z')},
\label{eq33}
\end{equation}
while $H(z)$ is calculated by the quintessence model (i.e. Eqs.(\ref{eq5}) and (\ref{eq6})).

\begin{figure}
  \centering
  \includegraphics[width=0.8\textwidth]{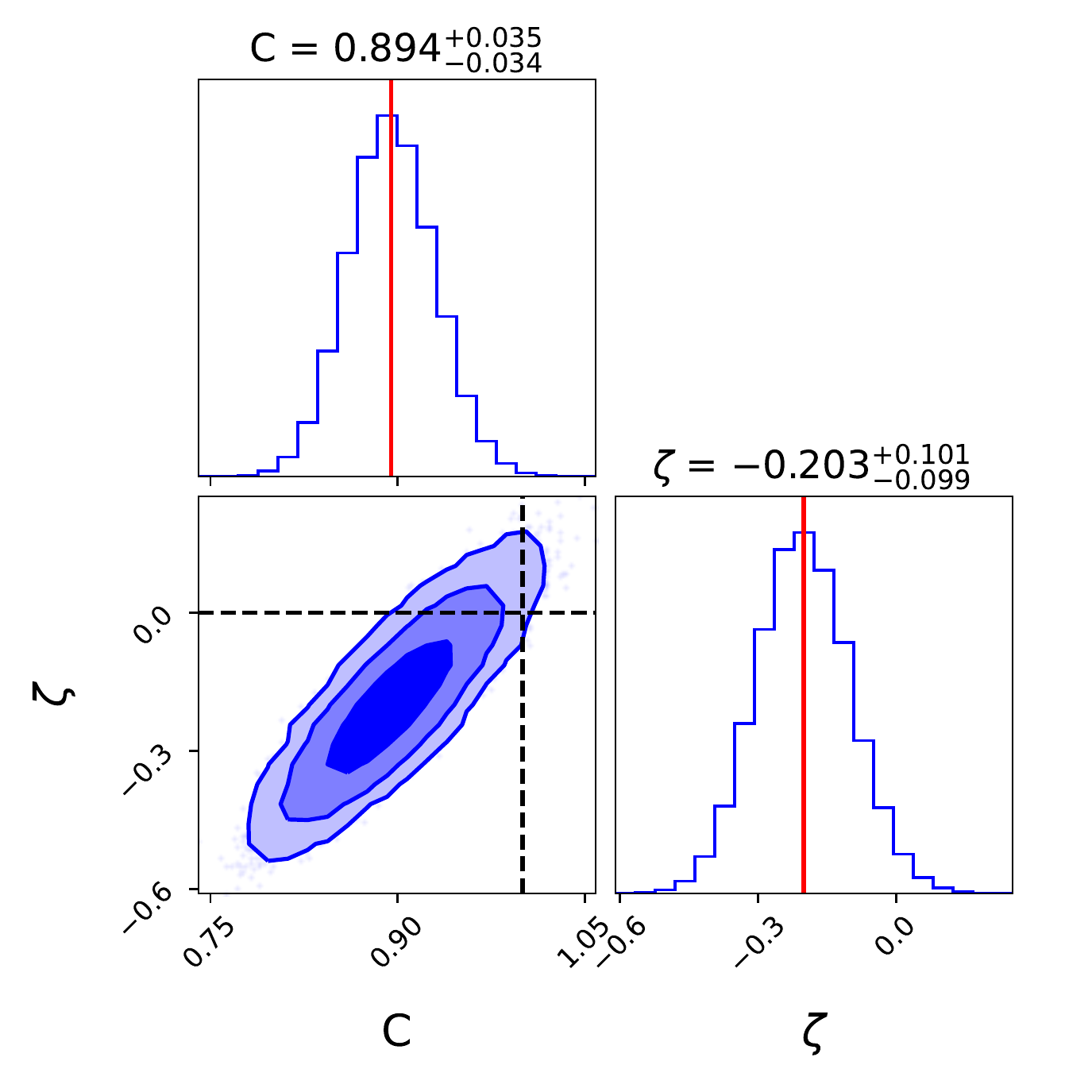}
  \caption{Contours of $1\sigma$, $2\sigma$ and $3\sigma$ on the $\zeta$ - $C$ plane and the corresponding 1-D marginalized likelihood distributions for the Planck clusters.}
  \label{fig2}
\end{figure}

\begin{figure}
  \centering
  \includegraphics[width=0.8\textwidth]{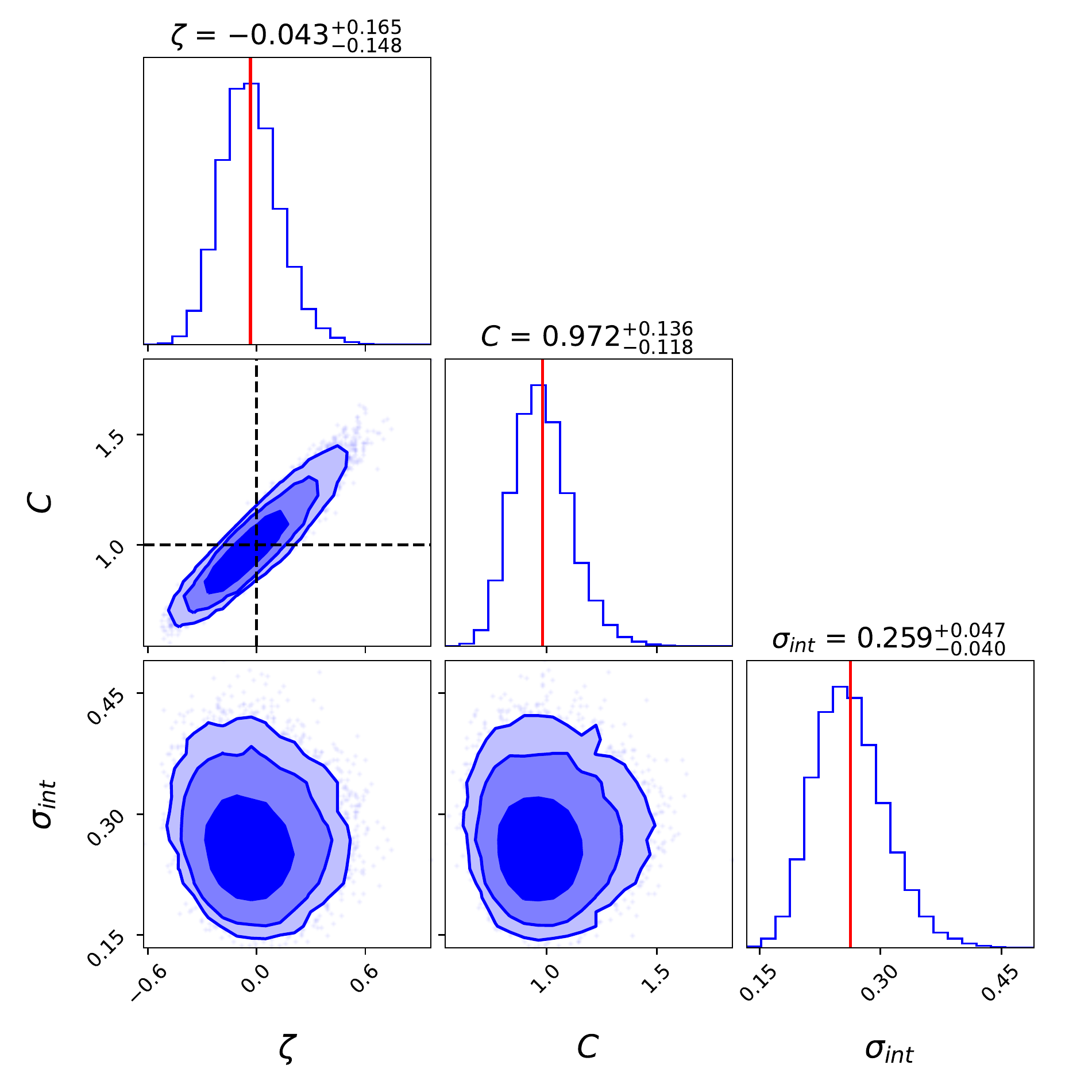}
  \caption{Contours of $1\sigma$, $2\sigma$ and $3\sigma$ on the 2-D parameter planes and the corresponding 1-D marginalized likelihood distributions for the SPT clusters. The red line represents the mean value of each parameter. The dashed lines in the $\zeta-C$ plane mark $\zeta = 0$ and $C=1$, which represents $\alpha = \alpha_0$ and the isothermality
assumption, respectively.}
  \label{fig3}
\end{figure}

We maximize the likelihood using the emcee MCMC sampler \citep{Foreman2013}. The results on the Planck cluster sample and those on the SPT cluster sample are given in Fig.\ref{fig2} and Fig.\ref{fig3}, respectively, which display the $68\%$, $95\%$, and $99\%$ confidence level plots along with the marginalized one-dimensional distribution of each parameter. We obtain $\zeta=-0.203^{+0.101}_{-0.099}$, $C = 0.894^{+0.035}_{-0.034}$ from the Planck sample and $\zeta = -0.043^{+0.165}_{-0.148}$, $C = 0.972^{+0.136}_{-0.118}$, $\sigma_{int} = 0.259^{+0.047}_{-0.040}$ from the SPT sample. We find from Fig.\ref{fig2} that the point ($\zeta=0,C=1$) is located within the $3\sigma$ contour for the Planck sample. Therefore, our results indicate that there is no significant evidence for nonzero $\zeta$, implying no time variation in $\alpha$. Furthermore, since $C=1$ holds within $3\sigma$ and $1\sigma$ for the Planck sample and the SPT sample respectively, we conclude that the isothermality assumption of the temperature profile is applicable quite well for both the Planck cluster sample and the SPT cluster sample. It seems that the intrinsic uncertainty of the SPT data ($\sigma_{int}=0.259$) is heavier than that of the Planck data ($\sigma_{int}=0.17$). Considering they stem from different analysis method, this conclusion needs further validation.

\section{Conclusions}
We explore the possible time variation of the fine structure constant $\alpha$ based on the scaling relation $Y_{SZ}D^2_A/C_{XYZ}Y_X$ calculated from galaxy cluster measurements. Following \citet{Colaco2019}, this ratio depends on the fine structure constant through the Thompson cross-section and is also affected by the violation of cosmic distance duality relation through the cluster gas mass. Instead of the frequently used runaway dilaton models, we use the quintessence model to provide the theoretical mechanism of generating time-varying $\alpha$. Different from the runaway dilaton model, for which the redshift dependence of the dilaton field is explicitly approximated by $\phi(z)=1-\gamma\ln(1+z)$ at low and intermediate redshifts, the dynamics of the quintessence scalar field is given by an evolution procedure described using Eqs.(\ref{eq5}) and (\ref{eq6}). The resulting equality ${Y_{SZ}D_A^2/Y_XC_{XSZ}} = C(1-\zeta(\phi-\phi_0))^{-3}$ is derived to link the quintessence field with the cluster observations.

We use the data of two cluster samples to constrain the variation of the fine structure constant. One sample is the observations of 61 galaxy clusters reported by the Planck collaboration, and another is the 58 galaxy clusters selected by the SPT-SZ observation. The X-ray counterpart (i.e. $Y_X$) of both the samples has been observed by the XMM-Newton observations. We give different treatments to the intrinsic uncertainty from the two data sets. For the Planck sample, we set $\sigma_{int}=0.17$ directly, while for the SPT sample $\sigma_{int}$ is taken as a free parameter and the determination of its best value is fused in the optimization procedure. In this way, the intrinsic uncertainty $\sigma_{int}$ for the SPT sample depends on the dispersion of angular diameter distance $D_A$ and thus is affected by the uncertainties of the cosmological parameters (i.e., $H_0$ and $\Omega_m$) indirectly. Our analyses show no significant evidence for the fine structure constant $\alpha$ varying with redshift, consistent with previous galaxy cluster-based results \citep{Galli2013,Colaco2019,Bora2021a,Bora2021b}. The constant $C$ was found to approach unity sufficiently, indicating the isothermal temperature profile is universal for describing the galaxy clusters.

 The number of galaxy clusters available is mainly limited by the current X-ray observations. For example, among the 516 clusters detected by the SPT-SZ survey, only 73 clusters have corresponding XMM-Newton X-ray observations. The eROSITA satellite, which has been launched in 2019, will perform the first imaging all-sky survey in the medium energy X-ray range and detect 50-100 thousand galaxy clusters \citep{Hofmann2017}. More detailed and richer observations in future may let us check whether the conclusion drawn in this paper is still valid in different redshift domains or not. We expect the problem of $\alpha$ variation would get a better answer with enlarging galaxy cluster samples.

\begin{acknowledgments}
We thank the anonymous referee for helpful and constructive feedback. This work was supported by the National Key R$\&$D Program of China (2017YFA0402600), the Natural Science Foundation of Shandong Province, China (Grant NO.ZR2019MA059), and the National Natural Science Foundation of China (grant No.11929301). ZZ is supported in part by NASA grant 15-WFIRST15-0008, Cosmology with the High Latitude Survey Roman Science Investigation Team (SIT). KB acknowledges Department of Science and Technology, Government of India for providing the financial support under DST-INSPIRE Fellowship program.
\end{acknowledgments}

\begin{figure}
  \centering
  \includegraphics[width=0.8\textwidth]{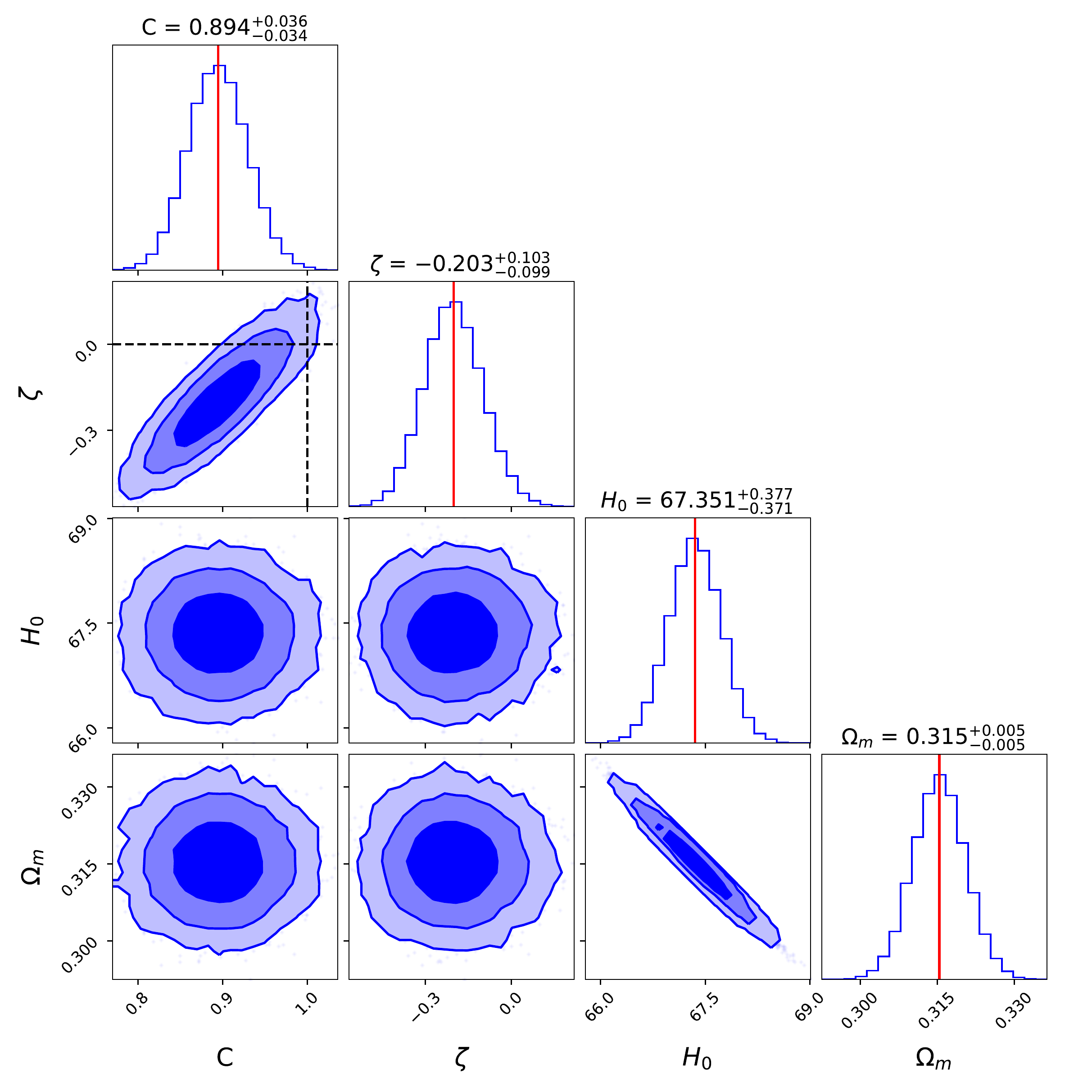}
  \caption{Contours of $1\sigma$, $2\sigma$ and $3\sigma$ on the 2-parameter planes and the corresponding 1-D marginalized likelihood distributions for the Planck clusters.}
  \label{fig4}
\end{figure}

\begin{figure}
  \centering
  \includegraphics[width=0.8\textwidth]{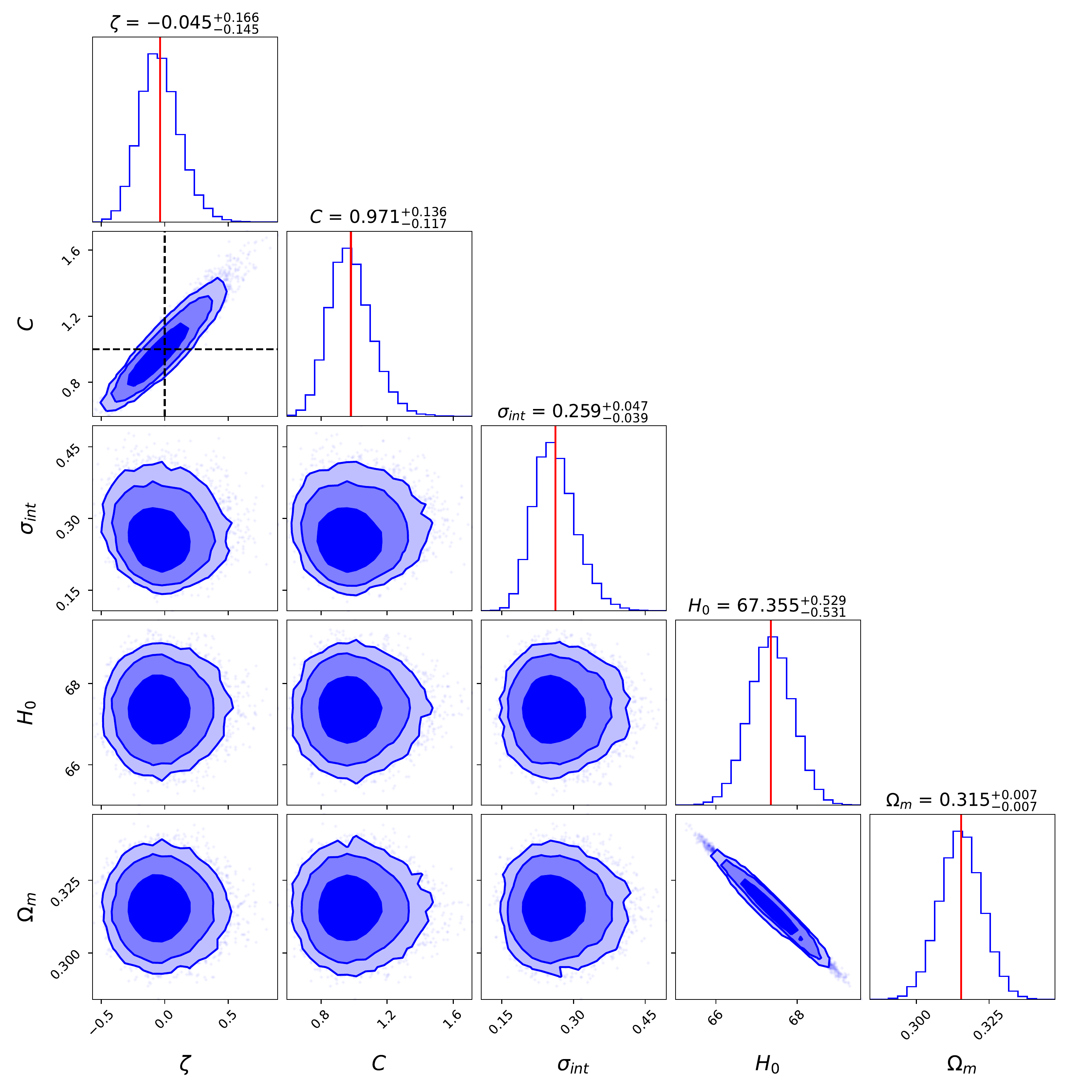}
  \caption{Contours of $1\sigma$, $2\sigma$ and $3\sigma$ on the 2-parameter planes and the corresponding 1-D marginalized likelihood distributions for the SPT clusters.}
  \label{fig5}
\end{figure}

\appendix
To demonstrate that the constraints on the time variation of $\alpha$ is insensitive to the cosmological parameters, we repeat the analysis for both data-sets treating $H_0$ and $\Omega_m$ as free parameters. We assume Gaussian bivariate prior distributions on $H_0$ and $\Omega_m$ with the following expression
\begin{equation}
p^{priors}(x)=\frac{1}{2\pi\sqrt{|\Sigma|}}\exp\left(-\frac{1}{2}(x-d)\Sigma^{-1}(x-d)\right),
\label{eq34}
\end{equation}
where $x=\{H_0,\Omega_m\}$, and their mean values $d=\{67.36,0.3153\}$ and covariance matrix $\Sigma=\begin{pmatrix}0.287 & -3.89\times10^{-3} \\ -3.89\times10^{-3} & 5.37\times10^{-5} \\ \end{pmatrix}$ are derived from the CosmoMC chains of Planck 2018 TT,TE,EE+lowE+lensing~\citep{Chen2019}. We get $C=0.894^{+0.036}_{-0.034}$, $\zeta=-0.202^{+0.103}_{-0.101}$ for the Planck clusters (see Fig.\ref{fig4}) and $\zeta=-0.045^{+0.166}_{-0.145}$, $C=0.971^{+0.136}_{-0.117}$, $\sigma_{int}=0.259^{+0.047}_{-0.039}$ for the SPT clusters (see Fig.\ref{fig5}). One can see that compared to the results obtained with fixed $H_0$ and $\Omega_m$, the difference between the both analyses is insignificant.


\bibliography{refs}{}
\bibliographystyle{aasjournal}

\end{CJK*}
\end{document}